\documentclass[nohyper]{JHEP3}

\usepackage{amsfonts}
\usepackage[centertags]{amsmath}
\usepackage{amssymb}
\usepackage{epsfig}
\usepackage{subfigure}
\usepackage[section]{placeins}

\providecommand{\mislash}[1]{#1 \mspace{-10.0mu} \slash}

\providecommand{\proarrow}[0]{\rightarrow}

\providecommand{\dif}[0]{\mathrm{d}}

\providecommand{\notll}[0]{\ll \mspace{-19.0mu} \slash \; \,}

\providecommand{\proname}[2]{#1 \proarrow #2}

\providecommand{\abs}[1]{\left\lvert #1 \right\rvert}
\providecommand{\deter}[1]{\left\lvert #1 \right\rvert}
\providecommand{\abst}[1]{\bigl\lvert #1 \bigr\rvert}

\providecommand{\miim}[1]{{\rm Im} \left[ #1 \right]}

\providecommand{\yde}[0]{Y_{\Delta_e}}
\providecommand{\ydm}[0]{Y_{\Delta_\mu}}
\providecommand{\ydt}[0]{Y_{\Delta_\tau}}
\providecommand{\gel}[0]{\gamma_e}
\providecommand{\gm}[0]{\gamma_\mu}
\providecommand{\gt}[0]{\gamma_\tau}
\providecommand{\gem}[0]{\gamma_{e \mu}}
\providecommand{\get}[0]{\gamma_{e \tau}}
\providecommand{\gmt}[0]{\gamma_{\mu \tau}}
\providecommand{\se}[0]{S_e}
\providecommand{\sm}[0]{S_\mu}
\providecommand{\st}[0]{S_\tau}
\providecommand{\gdlu}[0]{\gamma_{ \scriptscriptstyle{\Delta L=1}}}
\providecommand{\gfci}[0]{\gamma_{ \scriptscriptstyle{\text{FCI}}}}

\title{On fast CP violating interactions in leptogenesis}

\author{Chee Sheng Fong\\
  C.N. Yang Institute for Theoretical Physics\\
  State University of New York at Stony Brook\\
  Stony Brook, NY 11794-3840, USA,\\
  E-mail: \email{fong@insti.physics.sunysb.edu}}
\author{J. Racker\\
  Departament d'Estructura i Constituents de la Mat\`eria,
and Institut de Ci\`encies del Cosmos \\  
  Universitat de Barcelona,
  Diagonal 647, E-08028 Barcelona, Spain\\
  E-mail: \email{racker@ecm.ub.es}}

\abstract{We show that when the relevant CP violating interactions in leptogenesis are fast, the different matter density asymmetries are determined at each instant by a balance condition between the amount of asymmetry being created and destroyed. This fact allows to understand in a simple way many features of leptogenesis in the strong washout regime. In particular, we find some non-trivial effects of flavour changing interactions that conserve lepton number, which are specially relevant in models for leptogenesis that rely heavily on flavour effects.}
\preprint{%
YITP-SB-10-06 \\ ICCUB-10-028 }

\begin{document}
\section{Introduction}

Leptogenesis is one of the most attractive mechanisms to explain the
origin of the matter-antimatter asymmetry of the Universe~\cite{fukugita86}. This is so because it arises naturally in simple extensions of the standard model (SM) which
can also explain why the neutrino masses are so tiny. In this mechanism a lepton asymmetry is produced in the out of equilibrium decay of heavy Majorana neutrinos, which is then partially converted into a baryon asymmetry by non-perturbative sphaleron processes (see~\cite{davidson08} for a complete review). 

Among the three Sakharov conditions~\cite{sakharov67} for dynamically generating a baryon asymmetry, the one requiring out of equilibrium evolution is the least trivial regarding its quantitative aspects.  This condition is a consequence of CPT and unitarity, which imply that in a universe with fixed temperature and with all the particles having equilibrium phase space distributions, there can be no generation of asymmetry (see e.g.~\cite{weinberg79,kolb79,dolgov81}). However, if there are CP violating interactions mediated by particles which live for some time and the temperature of the universe is continuously changing, it is possible to generate a matter-antimatter asymmetry. This is so because the thermal bath in the universe is different when the mediating particles decay from what it was when the particles were created. Therefore the unitarity relations among CP violating processes, which hold at the level of probability amplitudes, cannot be extended to relations among the reaction densities -which are the quantities that drive the evolution of the density asymmetries- because they  also depend on the fluxes of the particles (see~\cite{roulet97} for a detailed explanation). In this work we will give some insights for the situations in which the CP violating interactions are very fast before decoupling and therefore the lifetime of the mediating particles is considerably smaller compared to the time scale set by the expansion rate of the universe. 
 
We will consider the evolution of the different matter density asymmetries as described by the classical Boltzmann equations (BE). In the equation for the lepton asymmetry, the role of the Yukawa interactions of the heavy neutrinos can be separated in two different type of terms. On one hand, the so called ``source terms'' involve the production of lepton asymmetry independently of the amount that is present in the Universe. On the other hand, there are ``washout terms'' proportional to the lepton asymmetry which tend to erase the asymmetry populating  the Universe at each instant. The main point raised in this work is that when the Yukawa interactions are much faster than the expansion rate of the Universe, at each instant the thermal bath has time to relax, in the sense that the amount of asymmetry being produced equals the amount being destroyed. This observation leads to a simple and accurate approximation for the evolution of the lepton asymmetry valid whenever the relevant Yukawa interactions are fast relative to the expansion rate and the lepton asymmetry is not changing abruptly (we will be more precise about this later). The approximation cannot describe the very last stage of leptogenesis when the asymmetry freezes, therefore it is not so worthful for obtaining accurate values of the final matter-antimatter asymmetry (which can be obtained solving numerically the BE or in many cases using analytical approximations~\cite{kolb90,buchmuller04,abada06II}). Notwithstanding it is useful to understand in a simple way many features of leptogenesis, in particular the dependence of the produced lepton asymmetry on the different parameters of the model. Furthermore, an advantage compared to the existing analytical approximations is that, being based on a simple physical argument, the approximation presented here can readily be generalized to include flavour effects and spectator processes, and its application to different types of models for leptogenesis is in many cases straightforward.  

The paper is organized as follows. In Sec.~2 we describe the main ideas and introduce the approximation, starting with the unflavoured case and then generalizing to the flavoured one. In Sec.~3 we apply the approximation to study the role of flavour changing interactions (FCI), i.e. interactions between different lepton flavours which conserve total lepton number, showing some non trivial effects and in Sec.~4 we summarize. 

\section{The strong washout balance approximation}
\label{sec:theaprox}
The idea developed here is very general, but for the sake of concreteness we will start working in the framework of the type I non-supersymmetric model for leptogenesis. In this model the particle contents is that of the Standard Model (SM) plus a number of heavy Majorana neutrinos $N_i$. The $N_i$ are SM singlets and only interact with the Higgs and the SU(2) lepton doublets through Yukawa interactions. Therefore, in a basis in which 
$N_i$ are mass eigenstates the Lagrangian reads 
\begin{equation}
\mathcal{L} = \mathcal{L}_{\text{SM}} + i \overline{N}_i
  \mislash{\partial} N_i - M_i \overline{N}_i
  N_i 
- \lambda_{\alpha i}\,{\widetilde h}^\dag\, \overline{P_R N_i} \ell_\alpha 
- \lambda^*_{\alpha i } \overline{\ell}_\alpha P_R N_i {\widetilde h},
\label{eq:lag}
\end{equation}
where $\alpha,i$ are family indices ($\alpha=e, \mu, \tau$ and 
$i=1,2,3, \dots$), $\ell_\alpha$ are the leptonic $SU(2)$ doublets, 
$h=(h^+,h^0)^T$ is the Higgs field ($\widetilde h =i\tau_2 h^*$, with $\tau_2$
Pauli's second matrix), and $P_{R,L}$ are the chirality projectors. 
Furthermore, we will mainly focus on hierarchical scenarios, for which the $N_1$ mass ($M_1$) is much smaller than the masses of the other heavy neutrinos and we also  assume that the lepton asymmetry generated during $N_{i>1}$ decays is not relevant. Hence, from now on the notation is simplified to $N \equiv N_1$ and $M \equiv M_1$.

Depending on the intensity of the Yukawa interactions, leptogenesis may occur in two very different regimes. The weak washout regime is defined by the condition $\Gamma_{N} \ll H(T=M)$, where $\Gamma_{N} = M (\lambda^\dag \lambda)_{11} / 8 \pi$ is the decay width of $N$ and $H$ is the Hubble rate. In this case the heavy neutrinos decay far out from equilibrium and the erasure of the asymmetry produced in the decay epoch is negligible. On the contrary, the strong washout regime [$\Gamma_{N} \gg H(T=M)$] is characterized by small departures from equilibrium and fast processes which tend to erase the lepton asymmetry. Alternatively, the conditions defining the weak and strong washout regimes can be expressed as $\tilde m \ll 10^{-3}$~eV and $\tilde m \gg 10^{-3}$~eV, respectively, where $\tilde m \equiv (\lambda^\dag \lambda)_{11} v^2 / M$ is the so called effective mass and $v=174$~GeV is the Higgs vacuum expectation value. 

Intuitively one can expect that when the CP violating interactions involving the heavy neutrinos are much faster than the expansion rate of the Universe (i.e. in the strong washout regime), the lepton asymmetry at each instant takes the value that enforces a perfect balance between the production and destruction rates of asymmetry. Actually, the requirements for this {\it balance condition} to rule the evolution of the lepton asymmetry are somewhat more restrictive. It does not hold at the initial stage of leptogenesis, because at the beginning the evolution is driven mainly by the source term. Also, when there is a change of sign in the lepton asymmetry, its evolution has nothing to do  with a balance between source and washout. Next we study when and how the balance condition applies to the case of unflavoured leptogenesis and then we generalize to the flavoured one.

\subsection{One flavour case}
\label{sec:unfalvoured}
The BE for the lepton asymmetry has the form 
\begin{equation}
\label{eq:b1f}
\frac{\dif Y_L}{\dif z} = S(z) - Y_L W(z) \; , 
\end{equation}     
where $z \equiv M/T$, $S(z)$ and $Y_L W(z)$ are the source and washout terms, respectively, and $Y_L$ is the lepton number density asymmetry normalized to the entropy density. The balance condition introduced above states that, at each instant $S(z) \approx Y_L W(z)$ and therefore, when it is valid, the lepton asymmetry takes the value 

\begin{equation}
\label{eq:swa1f}
Y_L(z) \approx \frac{S(z)}{W(z)} \; . 
\end{equation} 
We will call this approximate expression for the lepton density asymmetry the {\it strong washout balance approximation} (SWBA).

From the Eq.~\eqref{eq:b1f} it is clear that the balance condition is equivalent to $\tfrac{\dif Y_L}{\dif z} \approx 0$ and that the SWBA will be accurate as long as $\tfrac{\dif Y_L}{\dif z} \ll S(z), Y_L W(z)$. This helps to understand the picture in the regime well described by the SWBA: There are two relevant scales of time, a short one given by the rate of the CP violating interactions and a much longer one set by the expansion rate of the Universe\footnote{Actually, there is yet another important time scale associated with the gauge interactions that keep the massless particles (e.g. the leptons and the Higgs) in thermal equilibrium. This is typically much shorter than the other scales (see e.g.~\cite{kolb90}), as assumed in this work.}. At each instant there is some production of lepton asymmetry (given by the source term) and fast processes which tend to erase it. In the short time scale the production and erasure rates are constant and the lepton asymmetry takes the value that equilibrates the production and destruction processes.  The change in the lepton asymmetry is only due to the slow variation of the production and destruction rates driven by the temperature decrease of the expanding Universe. It is worth recalling that the -slow- expansion of the Universe is fundamental for generating an asymmetry. If the Universe were at a fixed temperature, there would be nothing like a source term and any initial asymmetry would be erased. Below we comment on an approximate expression for the source term that clearly shows its dependence on the expansion rate.

Considering the validity requirements for the SWBA, namely $\tfrac{\dif Y_L}{\dif z} \ll S(z), Y_L W(z)$, it is clear why the balance condition does not hold neither at the beginning of leptogenesis, nor when the lepton asymmetry changes sign. In the former situation, $Y_L$ is very small and hence $\tfrac{\dif Y_L}{\dif z} \approx S(z)$. In the latter, the source term changes sign, therefore $\abs{\tfrac{\dif Y_L}{\dif z}} = \abs{S(z)} + \abs{Y_L W(z)}$. Furthermore, at the end of leptogenesis when the CP violating interactions decouple, all the three terms $\tfrac{\dif Y_L}{\dif z}, S(z), Y_L W(z)$ become very small and the SWBA is expected to fail (clearly the evolution during freeze out is not given by a balance condition).   

In order to see how well the SWBA behaves, let us take first a simple example of unflavoured leptogenesis including only decays and inverse decays. The BE for this case are
\begin{equation}
\label{eq:ex}
\begin{split}
\frac{\dif Y_N}{\dif z} & = - \frac{1}{zHs} \left( \frac{Y_N}{Y_N^{eq}} - 1 \right) \gamma_D \; ,\\
\frac{\dif Y_{B-L}}{\dif z} & =  - \frac{1}{zHs} \left\{ \epsilon  \left( \frac{Y_N}{Y_N^{eq}} - 1 \right) \gamma_D + \frac{Y_{B-L}}{Y_\ell^{eq}} \frac{\gamma_D}{2} \right\} \; ,
\end{split}
\end{equation}
where $\epsilon$ parametrizes the amount of CP violation per neutrino decay, $Y_x \equiv n_x/s$ is the number density of the quantity $x$ normalized to the entropy density $s$, $Y_{x-y} \equiv Y_x - Y_y$, and the superscript $eq$ denotes an equilibrium density. In particular, $Y_L$, $Y_B$, and $Y_\ell^{eq}$ represent the lepton density asymmetry, baryon density asymmetry and equilibrium density of one relativistic degree of freedom, respectively, all of them normalized to the entropy density. Furthermore, $\gamma_D$ is the reaction density of decays, which is given by $\gamma_D(z)= n_N \tfrac{K_1(z)}{K_2(z)} \Gamma_N$, with $K_n$ the modified Bessel functions.

Neglecting the derivative of the $Y_{B-L}$ asymmetry\footnote{Here and in the following we work with the $B-L$ density asymmetry, or the $B/3-L_\alpha$ asymmetries for the flavoured case, because these quantities are conserved by the electroweak sphaleron processes and hence the BE are simpler. In any case, all the previous arguments equally apply to these asymmetries. Furthermore, when the electroweak sphalerons are fast, the baryon and lepton asymmetries are related to $Y_{B-L}$ by $Y_B = c Y_{B-L}$ and $Y_L = (c-1) Y_{B-L}$, respectively, with $c$ a constant that depends on the particle contents of the model~\cite{harvey90}.} leads to the SWBA,
\begin{equation}
Y_{B-L}(z) =  \frac{S(z)}{ \frac{1}{zHs} \frac{\gamma_D}{2 Y_\ell^{eq} }} \; , 
\label{eq:apr1}
\end{equation}
where $S(z)=-\tfrac{1}{zHs}  \epsilon  \big( \tfrac{Y_N}{Y_N^{eq}} - 1 \big) \gamma_D$ is the source term.

To use the SWBA as it stands in the above equation, it is necessary to solve the BE for the density of the heavy neutrinos because  $S(z)$ depends on $Y_N$. However, this can be avoided using a simple and accurate approximation for $S(z)$. To obtain it, note that the source term is equal to $\epsilon \frac{\dif Y_N}{\dif z}$ (which is also true if scatterings are included consistently~\cite{abada06II,nardi07II}). Therefore, once the density of heavy neutrinos begins to follow closely the equilibrium density, the source term can be approximated by  $\epsilon \frac{\dif Y_N^{eq}}{\dif z}$ and the SWBA reads
\begin{equation}
Y_{B-L}(z) =  \frac{\epsilon \frac{\dif Y_N^{eq}}{\dif z}}{ \frac{1}{zHs} \frac{\gamma_D}{2 Y_\ell^{eq} }} \; .
\label{eq:apr2}
\end{equation}

\begin{figure}[!htb]
\centerline{\protect\hbox{
\epsfig{file=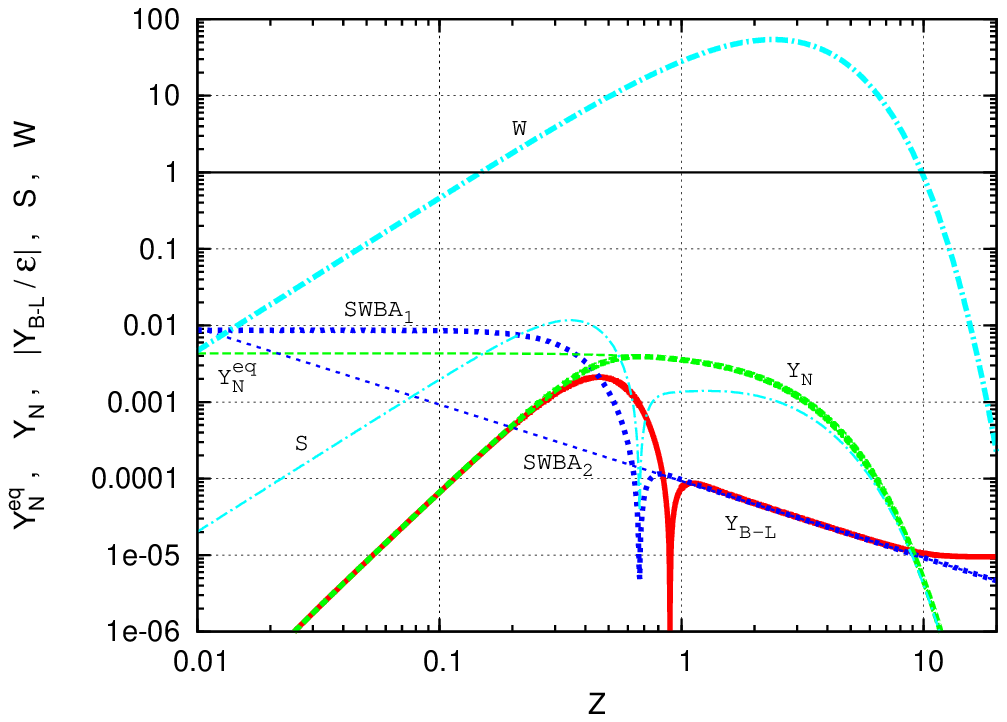 ,width=0.46\textheight,angle=270}}}
\caption[]{Densities and asymmetries in absolute value as a function of $z$. The thin dashed green curve is $Y_N^{eq}$, the thick dashed green curve is $Y_N$, the solid red curve is the exact absolute value of $Y_{B-L}$ normalized to $\epsilon$, the thick dotted blue curve (labeled $\text{SWBA}_1$) is the SWBA for $\abs{Y_{B-L}/\epsilon}$ using the exact expression for the source term, and the thin dotted blue curve (labeled $\text{SWBA}_2$) is the SWBA for $\abs{Y_{B-L}/\epsilon}$ approximating the source term to $ \epsilon \tfrac{\dif Y_N^{eq}}{\dif z}$. It is also shown the absolute value of the source term normalized to $\epsilon$, i.e. $\abs{\tfrac{\dif Y_N}{\dif z}}$ (thin dash dotted light blue curve) and  the inverse decay rate conveniently normalized, $\tfrac{1}{zHs} \frac{\gamma_D}{2 Y_\ell^{eq}}=W(z)$ (thick dash dotted light blue curve).} 
\label{fig:1}
\end{figure}

In Fig.~\ref{fig:1} we show the exact solution to the BE~\eqref{eq:ex} obtained by numerical integration, the SWBA using the exact expression for the source term [Eq.~\eqref{eq:apr1}], and the SWBA with the approximated source term [Eq.~\eqref{eq:apr2}]. For comparison the quantities $Y_N^{eq}, S(z),$ and $W(z) = \tfrac{1}{zHs} \frac{\gamma_D}{2 Y_\ell^{eq}}$ are also plotted. Note that $W(z)$ gives a measure of the ratio between the two time scales mentioned above. To give a clear example of the SWBA the effective mass has been set to $\tilde m = 0.1$~eV, which is a rather high value,  but we have checked that the accuracy of the approximation for effective masses one order of magnitude smaller is rather similar (it just works somewhat worse at the end of leptogenesis when the Yukawa interactions decouple). For this value of $\tilde m$ it is possible to have successful leptogenesis if the mass of the heavy neutrino is around $10^{13}$~GeV~(see e.g.~\cite{giudice04}), hence we have taken $M=10^{13}$~GeV. The figure illustrates that the SWBA works very well after the lepton asymmetry changes sign and until the asymmetry freezes at $z \sim 10$. 

We remark that the following things happen at $z \sim 10$: (I)~The washouts decouple, i.e. $W(z)$ becomes less than 1. (II)~The density of $N$ and the $B-L$ asymmetry (normalized to $\epsilon$) intersect, i.e. $\abs{Y_{B-L}} \approx \abs{\epsilon} Y_N$. The quantity $-\epsilon Y_N$ has a physical meaning, namely it is the total amount of $B-L$ asymmetry that can be produced in the decays of all the neutrinos present at a given moment (note that it is equal to the integral of the source term from that moment until $z \to \infty$). Therefore, at $z \sim 10$ the maximum amount of $B-L$ asymmetry that could be produced becomes less than the amount already present. (III)~The lepton asymmetry starts to freeze. This is a consequence of (I) and (II). (IV)~The SWBA begins to fail. Nevertheless it still works rather fine at this point. To see this, note that when $W(z)=1$ the SWBA gives $\abs{Y_{B-L}/\epsilon}= \abst{\tfrac{\dif Y_N^{eq}}{\dif z}} \simeq Y_N^{eq}$ (the last equality is valid for $z \gtrsim 5$), in agreement with (II).

From the figure it can also be seen that the SWBA given in Eqs.~\eqref{eq:apr1} and \eqref{eq:apr2} become equal after $z \sim 1$. This means that the approximation for the source term introduced above, namely $S(z) \approx \epsilon \frac{\dif Y_N^{eq}}{\dif z}$, is very accurate after the population of heavy neutrinos starts to be -almost- in equilibrium. Using this expression it is particularly  clear that the amount of asymmetry that can be generated (by the source term) in a given unit of time depends on the expansion rate of the universe, since this determines the rate of change of the temperature (or equivalently of $z$).

To conclude with this first example, we discuss how the SWBA can be used to estimate the dependence of the matter-antimatter asymmetry on the parameters of the model. Besides the CP asymmetry $\epsilon$, the only relevant parameter in this simple example is the effective mass $\tilde m$. From the SWBA given in Eq.~\eqref{eq:apr2} it is clear that $Y_{B-L}$ is inversely proportional to $\tilde m$ during the period ruled by the balance condition. If the temperature $T_{dec}$ at which the CP violating interactions decouple were independent of $\tilde m$, then the relation $Y_{B-L} \propto {\tilde m}^{-1}$ would also hold with good accuracy for the final value of the $B-L$ asymmetry. Actually $T_{dec}$ depends on $\tilde m$ (see~\cite{buchmuller04} for an analytical expression), but since the dependence is mild -logarithmic to be more precise- and $Y_{B-L}$ decreases only linearly with $z$, the departure from the ``$1/\tilde m$ law'' is not much. This gives an explanation to the known fact that the final $B-L$ asymmetry in the strong washout regime depends on $\tilde m$ as $1/{\tilde m}^{(1+a)}$, where $a$ has a mild dependence with $\tilde m$ (see for instance~\cite{buchmuller04}). For example, $a \sim 0.15$ for $\tilde m \sim 0.01$~eV and it gets smaller for larger values of $\tilde m$.
\FloatBarrier
\subsection{Flavoured case}
\label{sec:flavoured}
When one or more Yukawa interactions of the charged leptons are in equilibrium and have higher rates than the Yukawa interactions of the heavy neutrinos, the flavour structure of the evolution equations plays an important role~\cite{barbieri99,endoh03,abada06,nardi06,blanchet06,desimone06}. Choosing bases that diagonalize the density matrices of leptons and antileptons, the BE for the asymmetries can be written as
\begin{equation}
\frac{\dif Y_\Delta}{\dif z} = S - W \; Y_\Delta \; , 
\end{equation}
where $Y_\Delta$ and $S$ are the column matrices containing the $Y_{\Delta_\alpha}$ density asymmetries (with $\Delta_\alpha \equiv B/3-L_{\alpha}$) and the source terms $S_\alpha$, respectively ($\alpha = 1, \dots, n_f$ is a flavour index and $n_f$ is the number of relevant flavours). Furthermore $W$ is the $n_f \times n_f$ matrix containing the washout factors. Note that $S_\alpha$ is proportional to the flavoured CP asymmetry $\epsilon_\alpha \equiv (\Gamma_\alpha - \bar \Gamma_\alpha)/\sum_\beta (\Gamma_\beta + \bar \Gamma_\beta)$, where $\Gamma_\alpha \equiv \Gamma(\proname{N}{\ell_\alpha h})$ and $\bar \Gamma_\alpha \equiv \Gamma(\proname{N}{\bar \ell_\alpha \bar h})$ are the decay widths of $N$ into $\ell_\alpha h$ and $\bar \ell_\alpha \bar h$, respectively. If all the lepton flavours have fast CP violating interactions, the situation is similar to the one described in the unflavoured case, namely that at every instant the amount of CP asymmetry produced in each flavour equals its erasure, or equivalently $\frac{\dif Y_\Delta}{\dif z} \approx 0$. This leads to the SWBA 
\begin{equation}
\label{eq:fapr}
Y_\Delta \; \approx \; W^{-1} \; S, 
\end{equation}
that generalizes Eq.~\eqref{eq:swa1f} for flavoured leptogenesis.

\begin{figure}[!htb]
\centerline{\protect\hbox{
\epsfig{file=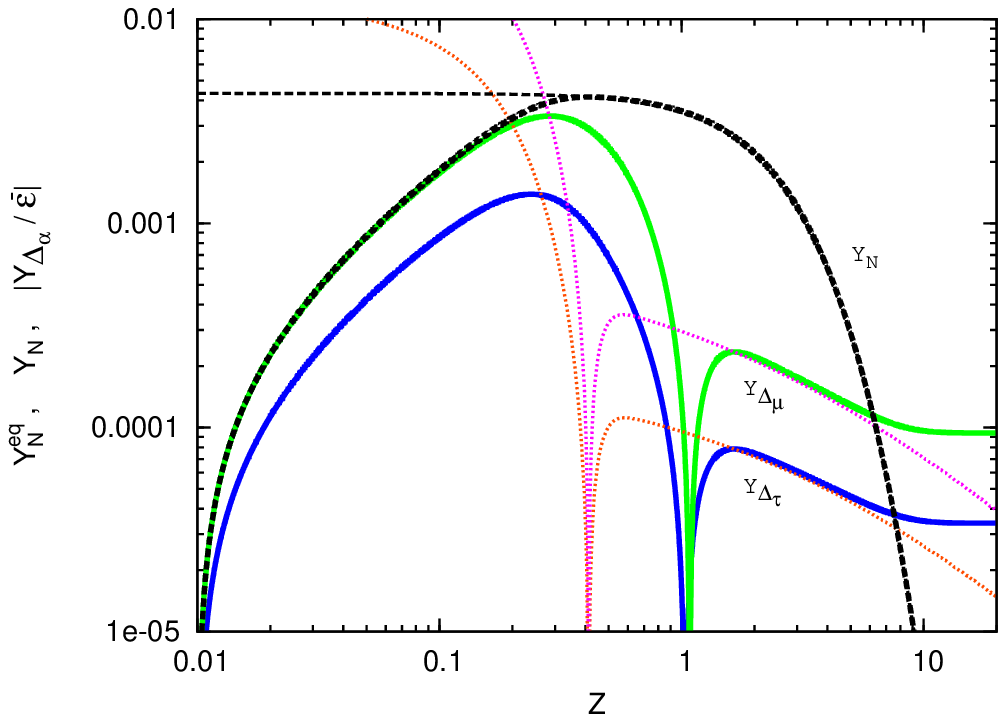 ,width=0.36\textheight,angle=270}}}
\caption[]{An example on the use of the SWBA for flavoured leptogenesis. The solid curves give the asymmetries $\abs{Y_{\Delta_\alpha}/\bar \epsilon}$  ($\alpha=\mu, \tau$) as a function of $z$ obtained by solving numerically the BE, while the dotted ones correspond to the SWBA represented symbolically in Eq.~\eqref{eq:fapr}. Also plotted are $Y_N$ (thick black dashed curve) and $Y_N^{eq}$ (thin black dashed curve).} 
\label{fig:2}
\end{figure}

Figure~\eqref{fig:2} shows an example on the use of the SWBA for a flavoured case. This time we have included decays, inverse decays and scatterings involving the top quark, but for simplicity we have ignored finite temperature corrections for the particle masses and couplings~\cite{giudice04}. The spectator processes active during leptogenesis are also included (see ~\cite{buchmuller01,nardi05} and \cite{nardi06}). The corresponding BE are the ones given in~\cite{nardi06} but modifying the source term to include the CP asymmetry in the scatterings as indicated in~\cite{nardi07II}. We have taken $M=10^{11}$~GeV and $\tilde m =0.05$~eV, in which case the Yukawa interaction of the $\tau$  -but not the one of the $\mu$- is faster than the expansion rate and the inverse decay of the heavy neutrino, therefore the dynamics can be described by two equations in flavour space~(\cite{blanchet06,desimone06,abada06,davidson08}). The flavour structure is given by the projectors $K_{\alpha} \equiv \Gamma_\alpha/\sum_\beta \Gamma_\beta$ and by the flavoured CP asymmetries. For the projectors we have chosen the values $K_e=0$, $K_{\mu}=0.2$, and $K_{\tau} = 1 - K_{e} - K_{\mu} = 0.8$. Furthermore, the flavoured CP asymmetries are taken to be proportional to some parameter $\bar \epsilon$,  $\epsilon_\mu = \bar \epsilon$ and $\epsilon_\tau = -0.5 \bar \epsilon$ (while $\epsilon_e=0$), therefore the $\Delta_\alpha$ asymmetries are simply proportional to $\bar \epsilon$ and in the figure we plot $\abs{Y_{\Delta_\alpha}/\bar \epsilon}$ (note that the total CP asymmetry $\epsilon \equiv \sum_\alpha \epsilon_\alpha = 0.5 \bar \epsilon$). From the figure it is clear that in the strong washout regime, after the $Y_{\Delta_{\alpha}}$ change sign and until they begin to freeze, the balance condition rules the evolution of the asymmetries also in flavoured cases. Also note that the ratios between the flavoured Yukawa interaction rates of the heavy neutrinos and the expansion rate of the universe can be read from the $W(z)$ curve in Fig.~\eqref{fig:1} weighted by the corresponding flavour projector and an overall 1/2 factor due to the different value of the effective mass used in this example. 
\section{Application to flavour changing interactions}
In this section we will apply the SWBA to study the role of FCI. This type of interactions is important, for example, when the Yukawa couplings of the heavy neutrinos are large, so that the processes mediated by virtual heavy neutrinos are fast, and also, as remarked in~\cite{aristizabal09}, in supersymmetric scenarios due to off diagonal soft breaking slepton masses. They can potentially enforce the equality of the lepton flavour density asymmetries, inhibiting large enhancements of the final baryon asymmetry due to flavour effects. Their effect could be even more drastic in models that have a null total CP asymmetry and rely completely on flavour effects (see e.g.~\cite{aristizabal09b,gonzalezgarcia09,antusch09}). This class of models cannot produce any baryon asymmetry if all the lepton flavour asymmetries are equal. Next we will argue that having fast FCI is not a sufficient condition for the flavour asymmetries to become equal and, in particular, that their effect in the models with null total CP asymmetry is less harmful than what could seem a priori.  

We introduce the notation $\gamma_\alpha$ for the sum of the interaction rates that produce $\abs{\Delta L} = \abs{\Delta L_\alpha} = 1$ and $\gamma_{\alpha \beta}$ for the sum of the ones that conserve $L$ but violate $L_\alpha$ and $L_\beta$ by one unit, all of them normalized in such a way that the BE can be written simply as\footnote{Spectator processes can easily be included in the analysis, but for simplicity we will neglect them.}
\begin{equation}
\label{eq:bfe}
\begin{array}{ccccccccc}
\displaystyle
\frac{\dif \yde}{\dif z} &=& \se &-& \gel \yde &-& \gem \left( \yde - \ydm \right) &-& \get \left( \yde - \ydt \right) \; ,  \\ &&&&&&&& \\
\displaystyle
\frac{\dif \ydm}{\dif z} &=& \sm &-& \gm \ydm &-& \gem \left( \ydm - \yde \right) &-& \gmt \left( \ydm - \ydt \right) \; , \\ &&&&&&&& \\
\displaystyle
\frac{\dif \ydt}{\dif z} &=& \st &-& \gt \ydt &-& \get \left( \ydt - \yde \right) &-& \gmt \left( \ydt - \ydm \right) \;.  
\end{array}
\end{equation}
When all the lepton flavours have fast CP violating interactions, the values of the $Y_{\Delta_\alpha}$ asymmetries are determined by the condition that all the derivatives $\frac{\dif Y_{\Delta_\alpha}}{\dif z}$ be negligible. The solution to the linear system of equations that results after setting the derivatives of the $Y_{\Delta_\alpha}$ asymmetries to zero is
\begin{equation}
\label{eq:fci}
\begin{split}
\yde =  & \; \left[ \gm \gt \se + \gem \gt (\se + \sm) + \get \gm (\se + \st) + \gmt (\gm + \gt) \se \right. \\ &\left. \;\; + (\gem \gmt + \get \gmt + \gem \get) (\se + \sm + \st) \right] \; / \; \deter{W} \; ,\\ & \\
\ydm =  & \; \left[ \gt \gel \sm + \gmt \gel (\sm + \st) + \gem \gt (\sm + \se) + \get (\gt + \gel) \sm \right. \\ &\left. \;\; + (\gmt \get + \gem \get + \gmt \gem) (\sm + \st + \se) \right] \; / \; \deter{W} \; ,\\ & \\
\ydt =  & \; \left[ \gel \gm \st + \get \gm (\st + \se) + \gmt \gel (\st + \sm) + \gem (\gel + \gm) \st \right. \\ &\left. \;\; + (\get \gem + \gmt \gem + \get \gmt) (\st + \se + \sm) \right] \; / \; \deter{W} \; ,
\end{split}
\end{equation}
where $\deter{W}$ is the determinant of the washout matrix, 
\begin{equation}
\begin{split}
 \; \deter{W} = & \gel \gm \gt + \gmt \gel (\gm + \gt) + \get \gm (\gel + \gt) + \gem \gt (\gel + \gm) \\
& + (\gem \get + \gem \gmt + \get \gmt) (\gel + \gm + \gt) \; .
\end{split}
\end{equation}

Each $\Delta_\alpha$ asymmetry in the above equations is given by a sum of terms which contain products of two interaction rates and one source factor, all of them divided by the common factor $\abs{W}$. Some terms involve the product of two $\Delta L=1$ interaction rates, others have the product of one $\Delta L=1$ interaction rate with one FCI rate, and finally there are terms with products of two FCI rates. The sum of this last type of terms is the same for every $\Delta_\alpha$ asymmetry, therefore all the $Y_{\Delta_\alpha}$ ($\alpha=e, \mu, \tau$) will be equal in the case that these terms dominate over the others. But for this to happen it is not enough to have fast FCI rates (i.e. faster than the expansion rate). Basically two more conditions must be satisfied. In order to state them in a simple way, let us assume that all the FCI rates have the same order of magnitude, with their average value denoted by $\gfci(z)$. In addition, all the $\Delta L=1$ interaction rates are also taken to be similar in  magnitude, with an average value equal to $\gdlu(z)$. Then, the aforementioned conditions are
\begin{itemize}
\item[(I)] $\gfci(z) \gg \gdlu(z)$ ,
\item[(II)] $\sum_\alpha S_\alpha \notll \, 1$ .
\end{itemize}
In particular, the condition (II) for ensuring equality of all the lepton flavour asymmetries is not trivial a priori, but follows rather directly from the analysis based on the SWBA. Since the models with null total CP asymmetry verify $\sum_\alpha S_\alpha = 0$, the asymmetries $Y_{\Delta_\alpha}$ will not be in general equal in these cases, even under the presence of very fast FCI (this holds during the period ruled by the balance condition but not after the $\Delta L=1$ interactions decouple, as will be further discussed in the example given below).

From the Eqs.~\eqref{eq:fci} it is also possible to estimate the dependence of the density asymmetries on the intensity of the FCI. We take $\gfci \propto {\tilde m_{od}}^4$, where ${\tilde m_{od}}^4$ parametrizes the intensity of the FCI (the reason for this notation will become clear below). For the case $\epsilon \approx 0$ the Eqs.~\eqref{eq:fci} indicate that if $\gfci \gg \gdlu$ before the CP violating interactions decouple, then $\abs{Y_{\Delta \alpha}} \propto 1/{\tilde m_{od}}^4$. Note that this statement, as well as the conditions (I) and (II) for having equal flavour asymmetries, hold for generic flavour structures of the Yukawa couplings, while special cases like the presence of null projectors or some projectors being equal have to be treated separately.

To show a detailed example we consider the case of soft leptogenesis~\cite{grossman03,dambrosio03}, taking into account the full flavour structure~\cite{fong08} and including the FCI induced by off diagonal soft supersymmetry breaking masses for the slepton doublets~\cite{aristizabal09}. A difference between the example given here and common analysis in the literature is that, in order to get a null total CP asymmetry, we do not assume the trilinear couplings ``$A$" of the soft breaking terms $A_{\alpha 1} \lambda_{\alpha 1} \tilde N \tilde \ell_\alpha h$ to be flavour independent. Hence the flavoured CP asymmetries in the resonant condition can be written  as $\epsilon_\alpha(T)= K_{\alpha} \miim{A_{\alpha 1}} \Delta_{BF}(T) / M$ (for one generation of heavy neutrinos), where $\Delta_{BF}(T)$ is a thermal factor that goes to 0 when $T \to 0$~\cite{dambrosio03}. For the example we have chosen $K_{e}=0.1, K_{\mu}=0.2,$ and $K_{\tau}=0.7$, whereas $\miim{A_{\alpha 1}} = \abs{A} \sin \phi_\alpha$, with $\abs{A}=10^4$~GeV, $\phi_{e}=\pi/2, \phi_{\mu}=\pi/4,$ and $\phi_{\tau}$ equal to the value that makes $\epsilon = \sum_{\alpha} \epsilon_\alpha=0$. Furthermore, we have calculated and included the rates of the FCI induced by the off diagonal soft breaking slepton masses, which are taken to be all equal to a parameter called $\tilde m_{od}$ (see~\cite{aristizabal09} for a detailed description), so that $\gfci \propto {\tilde m_{od}}^4$. The FCI rates also depend on $\tan \beta$, taken here to be equal to 30. For more details on the evolution equations and parameters involved the reader is referred to~\cite{fong08}.

\begin{figure}[!htb]
\centerline{\protect\hbox{
\epsfig{file=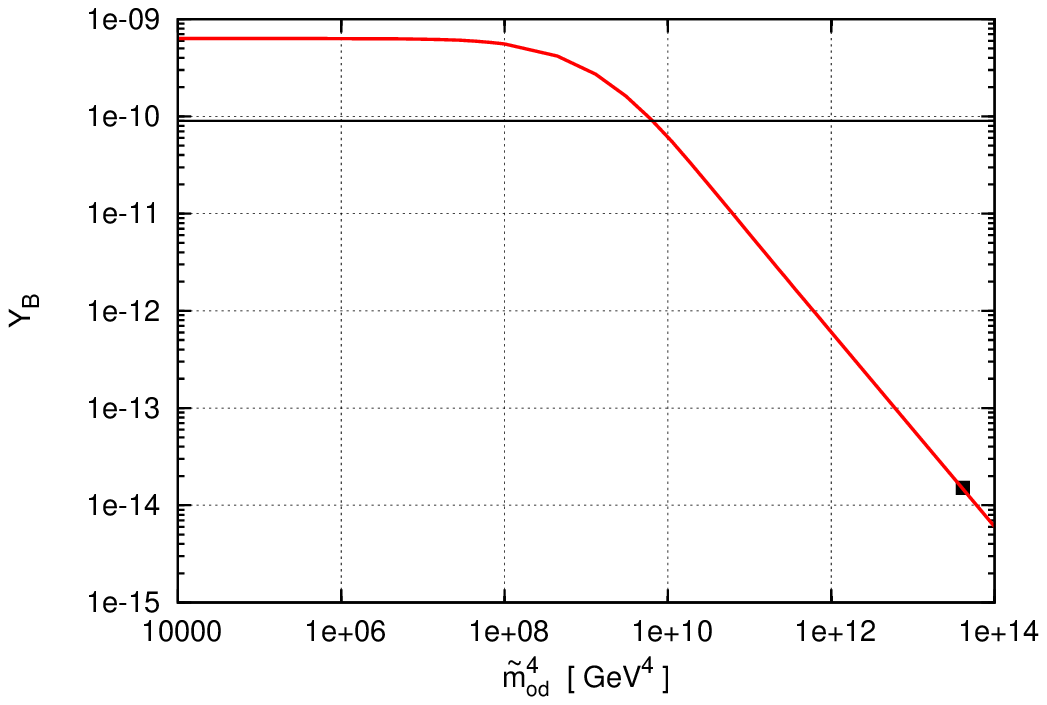 ,width=0.36\textheight,angle=270}}}
\caption[]{The final baryon asymmetry as a function of ${\tilde m_{od}}^4$ for the example described in the text. The horizontal line corresponds to the cosmological asymmetry and the small square is a reference for Fig.~\eqref{fig:4}.} 
\label{fig:3}
\end{figure}

\begin{figure}[!htb]
\centering
\subfigure[]{\label{subfig:4a}
\protect\hbox{
\epsfig{file=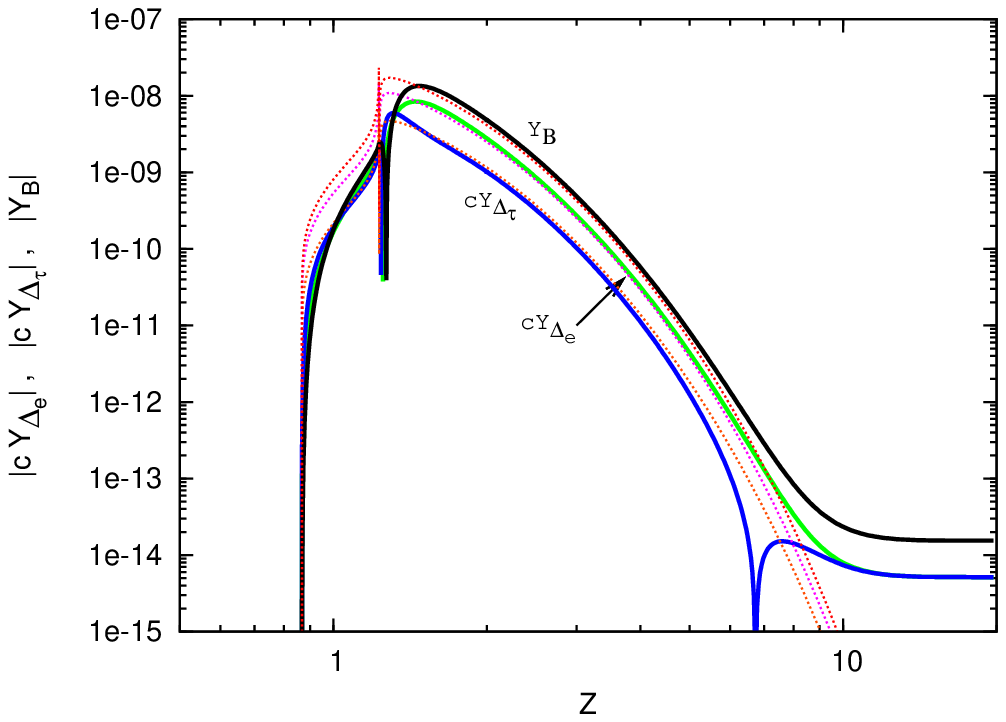,width=0.36\textheight,angle=270}}} 

\subfigure[]{\label{subfig:4b}
\protect\hbox{
\epsfig{file=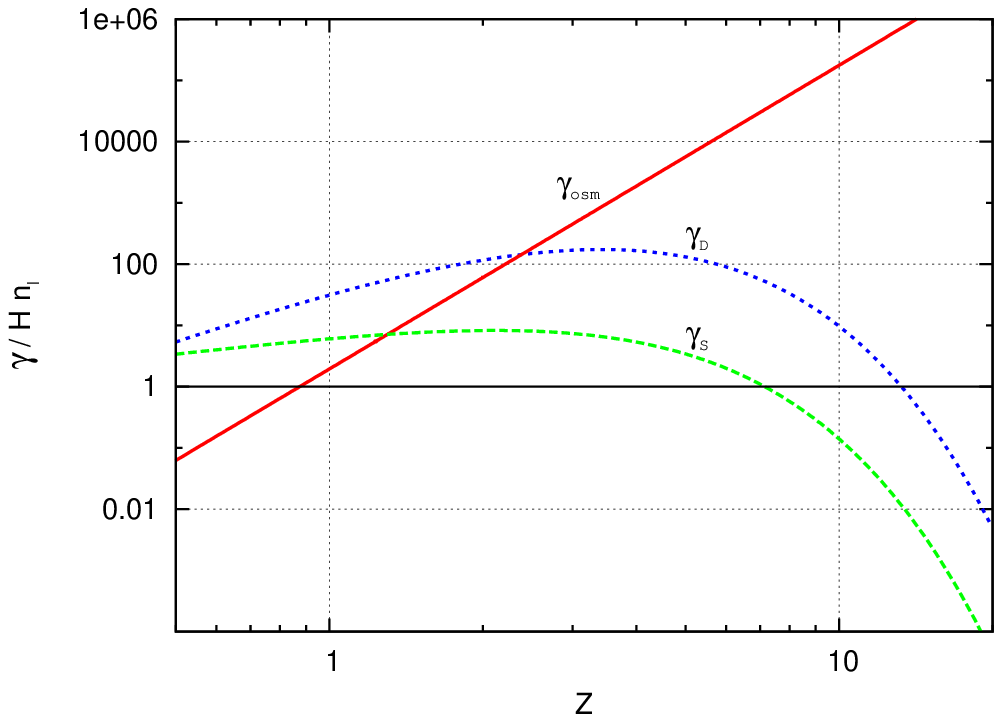,width=0.36\textheight,angle=270}}} 
\caption[]{

The asymmetries and rates as a function of $z$, corresponding to the point signaled with a small square in Fig.~\eqref{fig:3}. (a) The asymmetries $c Y_{\Delta_e}, c Y_{\Delta_\tau},$ and $Y_B= \sum_\alpha c Y_{\Delta_\alpha}$ as a function of $z$, where $c=8/23$ is the conversion factor between the $B-L$ and the $B$ asymmetries due to the electroweak sphaleron processes. The solid curves give the asymmetries  obtained by solving numerically the BE, while the dotted ones correspond to the SWBA. The asymmetry $Y_{\Delta \mu}$ is approximately equal to $Y_{\Delta_e}$ and hence it has not been plotted. (b) The reaction densities of decays summed over all flavours ($\gamma_D$), CP violating scatterings involving the top quark also summed over all flavours ($\gamma_s$), and FCI induced by off diagonal soft breaking slepton masses averaged over all flavours ($\gamma_{osm}$), as a function of $z$. All  the reaction densities have been normalized to $H n_l$, where $n_l$ denotes the equilibrium density of one relativistic degree of freedom.}
\label{fig:4}
\end{figure}

Figures~\eqref{fig:3} and \eqref{fig:4} contain the results of our analysis. In Figure~\eqref{fig:3} the final baryon asymmetry is plotted as a function of ${\tilde m_{od}}^4$ for a fixed value of the effective mass $\tilde m=0.2$~eV and $M_1=10^6$~GeV. It is clear that the behavior described above is verified. For small $\tilde m_{od}$ the asymmetry is independent of $\tilde m_{od}$, but when the FCI become fast enough, the final asymmetry satisfies $Y_{B-L} \propto 1/{\tilde m_{od}}^4$. The -approximate- value of the cosmological asymmetry is indicated with a horizontal line and it can be seen that in this example the cosmological asymmetry is achieved under FCI dominating over all the other rates, even though the total CP asymmetry is null. 

Figure~(4a) shows the evolution of the density asymmetries corresponding to the point signaled in Fig.~\eqref{fig:3}, which is representative of a situation with FCI much faster than the expansion rate and the $\Delta L=1$ interactions during most of the decay epoch [see Fig.~(4b)]. As can be seen the SWBA reproduces with reasonable accuracy the exact results. In particular the ``predicted'' non equality of all the flavour asymmetries during the leptogenesis era, even with very fast FCI, is verified. Note that the flavour asymmetries do equalize at the end of leptogenesis, but this happens when the $\Delta L=1$ interactions are decoupling and the $B-L$ asymmetry is freezing, so that $Y_{B-L}$ is not affected much. Instead, if the equality among the asymmetries had hold at earlier times, the total asymmetry would have been null, since $\epsilon=0$. 

From Figure~(4a) it can also be seen that in this case the SWBA ceases to be good somewhat before the freeze out, due to the late change of sign of $Y_{\Delta_\tau}$. Nevertheless, the dependence of $Y_{B-L}$ on a given parameter of the model does not change much between the period ruled by the balance condition and the end of leptogenesis, as long as $T_{dec}$ does not depend on this parameter (remember the discussion at the end of Sec.~2.1 and take $T_{dec}$ for the flavoured case equal to the temperature at which the largest CP violating interaction decouples). This is why the SWBA remains useful for obtaining the dependence of the final asymmetry on some parameters of the model, in particular on $\tilde m_{od}$ which does not change $T_{dec}$. On the contrary, the study of the dependence of $Y_{B-L}$ on $\tilde m$ in this example is more involved. While the SWBA holds, $Y_{B-L}$ is independent of $\tilde m$ when $\gfci \gg \gdlu$. But the final $B-L$ asymmetry gets some dependence on $\tilde m$ because $T_{dec}$ decreases with increasing $\tilde m$. Contrary to the simple case described in Sec.~2.1, here $Y_{B-L}$ is decreasing exponentially before freezing out\footnote{This can also be understood from the SWBA. The Eqs.~\eqref{eq:fci} for the case $\epsilon = 0$ and $\gfci \gg \gdlu$ roughly imply that $Y_{\Delta_\alpha}(z) \sim S(z) / \gfci(z)$, where $S(z)$ is some combination of the source terms. Since $S(z) \propto \tfrac{\dif Y_N^{eq}}{\dif z} \propto e^{-z}$ for $z \gtrsim 1$ and $\gfci$ is not exponentially suppressed, the exponential decrease of $Y_{B-L}$ with increasing $z$ follows.}, hence a small change in $T_{dec}$ produces a big change in the final value of $Y_{B-L}$. It can be shown that if the CP asymmetries $\epsilon_\alpha$ were independent of the temperature, then the final asymmetry would approximately satisfy $Y_{B-L} \propto 1/\tilde m$, as a result of the combination of the logarithmic dependence of $T_{dec}$ on $\tilde m$ with the exponential decrease of $Y_{B-L}$ before freeze out. But actually the CP asymmetries decrease with temperature, which adds an additional dependence of the final $B-L$ asymmetry on $\tilde m$. This could be studied more precisely using the SWBA together with a knowledge of  $T_{dec}=T_{dec}(\tilde m)$, however it is not the goal of this work to give more details on this point.

As a final remark, we want to refer to a model of leptogenesis with null total CP asymmetry presented in~\cite{gonzalezgarcia09}. Actually, a brief mention of the basic idea developed in this work was made there to gain understanding about the dependence of the baryon asymmetry on the different parameters. In this model, as well as other TeV scale models for leptogenesis (see e.g.~\cite{pilaftsis08}), the electroweak sphalerons decouple before the end of leptogenesis. Then, the baryon asymmetry often freezes in the regime well described by the balance condition and in these cases the results obtained from the SWBA are very accurate. 
\section{Conclusions}

It has been shown that when the CP violating interactions in leptogenesis are fast, the classical evolution of the lepton asymmetry is ruled by an equality condition between the amount of asymmetry that is being generated and erased. This fact can be used to obtain a simple approximation for the asymmetry, called here SWBA, which can generically be written as $Y_L=S(z)/W(z)$. Here $S(z)$ is the source term and $W(z)$ is a combination of interaction rates such that $W(z) Y_L$ is the whole washout term. This relation is also valid when different lepton flavours participate in the dynamics. In many models the source term can be obtained with high accuracy just from the rate of change of the heavy neutrinos equilibrium density induced by the expansion of the Universe. Therefore, the SWBA often does not require any integration of differential equations.

To be more precise, the balance condition between the generation and destruction of asymmetry holds as long as the CP violating interactions remain fast with the exception of the initial period in which the lepton asymmetry is starting to be generated or when it is changing sign. More importantly, the SWBA fails at the end of leptogenesis because the Yukawa interactions of the heavy neutrinos become slow, therefore in general it is not useful for obtaining accurate values of the final asymmetry. However, as explained in this work, the SWBA allows to understand in a simple and rather accurate way the dependence of the asymmetry on the parameters of the model. In particular, we have shown that it is useful for studying the models with null total CP asymmetry and fast FCI. In these cases flavour effects are mandatory to have successful leptogenesis and hence FCI could be potentially harmful. However, we have found that fast FCI  do not necessarily spoil baryogenesis, since the final asymmetry has a quite mild dependence on the intensity of these rates, namely $Y_B \propto 1/{\tilde m_{od}}^4$ (with the rate of the FCI being proportional to ${\tilde m_{od}}^4$). This relation is comparable to the dependence of the baryon asymmetry on the effective mass in simple models for leptogenesis which do not require flavour effects to be successful.
  

\section*{Acknowledgments}
We thank E.~Nardi and J.~Salvado for useful discussions and comments.
We are specially indebted to M.~C.~Gonzalez-Garcia and N.~Rius for discussions and careful reading of the manuscript.

This work is supported by Spanish  MICCIN  under grants 2007-66665-C02-01,
ACI2009-1038, and Consolider-Ingenio 2010  CUP (CSD2008-00037),
by CUR Generalitat de  Catalunya under project 2009SGR502,
and by USA-NSF grant PHY-0653342.
\providecommand{\href}[2]{#2}\begingroup\raggedright\endgroup



\begin{thebibliography}{10}

\bibitem{fukugita86}
M.~Fukugita and T.~Yanagida, {\it {Baryogenesis Without Grand Unification}},
  {\em Phys. Lett.} {\bf B174} (1986) 45.

\bibitem{davidson08}
S.~Davidson, E.~Nardi, and Y.~Nir, {\it {Leptogenesis}},  {\em Phys. Rept.}
  {\bf 466} (2008) 105--177,
  [\href{http://xxx.lanl.gov/abs/hep-ph/0802.2962}{{\tt hep-ph/0802.2962}}].

\bibitem{sakharov67}
A.~D. Sakharov, {\it {Violation of CP Invariance, c Asymmetry, and Baryon
  Asymmetry of the Universe}},  {\em Pisma Zh. Eksp. Teor. Fiz.} {\bf 5} (1967)
  32--35.

\bibitem{weinberg79}
S.~Weinberg, {\it {Cosmological Production of Baryons}},  {\em Phys. Rev.
  Lett.} {\bf 42} (1979) 850--853.

\bibitem{kolb79}
E.~W. Kolb and S.~Wolfram, {\it {Baryon Number Generation in the Early
  Universe}},  {\em Nucl. Phys.} {\bf B172} (1980) 224.

\bibitem{dolgov81}
A.~D. Dolgov and Y.~B. Zeldovich, {\it {Cosmology and Elementary Particles}},
  {\em Rev. Mod. Phys.} {\bf 53} (1981) 1--41.

\bibitem{roulet97}
E.~Roulet, L.~Covi, and F.~Vissani, {\it {On the CP asymmetries in Majorana
  neutrino decays}},  {\em Phys. Lett.} {\bf B424} (1998) 101--105,
  [\href{http://xxx.lanl.gov/abs/hep-ph/9712468}{{\tt hep-ph/9712468}}].

\bibitem{kolb90}
E.~W. Kolb and M.~S. Turner, {\it {The Early universe}},  {\em Front. Phys.}
  {\bf 69} (1990) 1--547.

\bibitem{buchmuller04}
W.~Buchm{\"u}ller, P.~Di~Bari, and M.~Pl{\"u}macher, {\it {Leptogenesis for
  pedestrians}},  {\em Ann. Phys.} {\bf 315} (2005) 305,
  [\href{http://xxx.lanl.gov/abs/hep-ph/0401240}{{\tt hep-ph/0401240}}].

\bibitem{abada06II}
A.~Abada, S.~Davidson, A.~Ibarra, F.~Josse-Michaux, M.~Losada, and A.~Riotto,
  {\it {Flavour matters in leptogenesis}},  {\em JHEP} {\bf 09} (2006) 010,
  [\href{http://xxx.lanl.gov/abs/hep-ph/0605281}{{\tt hep-ph/0605281}}].

\bibitem{harvey90}
J.~A. Harvey and M.~S. Turner, {\it {Cosmological baryon and lepton number in
  the presence of electroweak fermion number violation}},  {\em Phys. Rev.}
  {\bf D42} (1990) 3344--3349.

\bibitem{nardi07II}
E.~Nardi, J.~Racker, and E.~Roulet, {\it {CP violation in scatterings, three
  body processes and the Boltzmann equations for leptogenesis}},  {\em JHEP}
  {\bf 09} (2007) 090, [\href{http://xxx.lanl.gov/abs/hep-ph/0707.0378}{{\tt
  hep-ph/0707.0378}}].

\bibitem{giudice04}
G.~F. Giudice, A.~$\text{Notari}$, M.~Raidal, A.~Riotto, and A.~Strumia, {\it
  {Towards a complete theory of thermal leptogenesis in the SM and MSSM}},
  {\em Nucl. Phys.} {\bf B685} (2004) 89,
  [\href{http://xxx.lanl.gov/abs/hep-ph/0310123}{{\tt hep-ph/0310123}}].

\bibitem{barbieri99}
R.~Barbieri, P.~Creminelli, A.~Strumia, and N.~Tetradis, {\it {Baryogenesis
  through leptogenesis}},  {\em Nucl. Phys.} {\bf B575} (2000) 61--77,
  [\href{http://xxx.lanl.gov/abs/hep-ph/9911315}{{\tt hep-ph/9911315}}].

\bibitem{endoh03}
T.~Endoh, T.~Morozumi, and Z.~Xiong, {\it {Primordial lepton family asymmetries
  in seesaw model}},  {\em Prog. Theor. Phys.} {\bf 111} (2004) 123--149,
  [\href{http://xxx.lanl.gov/abs/hep-ph/0308276}{{\tt hep-ph/0308276}}].

\bibitem{abada06}
A.~Abada, S.~Davidson, F.~X. Josse-Michaux, M.~Losada, and A.~Riotto, {\it
  {Flavour issues in leptogenesis}},  {\em JCAP} {\bf 0604} (2006) 004,
  [\href{http://xxx.lanl.gov/abs/hep-ph/0601083}{{\tt hep-ph/0601083}}].

\bibitem{nardi06}
E.~Nardi, Y.~Nir, E.~Roulet, and J.~Racker, {\it {The importance of flavor in
  leptogenesis}},  {\em JHEP} {\bf 01} (2006) 164,
  [\href{http://xxx.lanl.gov/abs/hep-ph/0601084}{{\tt hep-ph/0601084}}].

\bibitem{blanchet06}
S.~Blanchet, P.~Di~Bari, and G.~G. Raffelt, {\it {Quantum Zeno effect and the
  impact of flavor in leptogenesis}},  {\em JCAP} {\bf 0703} (2007) 012,
  [\href{http://xxx.lanl.gov/abs/hep-ph/0611337}{{\tt hep-ph/0611337}}].

\bibitem{desimone06}
A.~De~Simone and A.~Riotto, {\it {On the impact of flavour oscillations in
  leptogenesis}},  {\em JCAP} {\bf 0702} (2007) 005,
  [\href{http://xxx.lanl.gov/abs/hep-ph/0611357}{{\tt hep-ph/0611357}}].

\bibitem{buchmuller01}
W.~Buchm{\"u}ller and M.~Pl{\"u}macher, {\it {Spectator processes and
  baryogenesis}},  {\em Phys. Lett.} {\bf B511} (2001) 74,
  [\href{http://xxx.lanl.gov/abs/hep-ph/0104189}{{\tt hep-ph/0104189}}].

\bibitem{nardi05}
E.~Nardi, Y.~Nir, J.~Racker, and E.~Roulet, {\it {On Higgs and sphaleron
  effects during the leptogenesis era}},  {\em JHEP} {\bf 01} (2006) 068,
  [\href{http://xxx.lanl.gov/abs/hep-ph/0512052}{{\tt hep-ph/0512052}}].

\bibitem{aristizabal09}
D.~Aristizabal~Sierra, M.~Losada, and E.~Nardi, {\it {Lepton Flavor
  Equilibration and Leptogenesis}},  {\em JCAP} {\bf 0912} (2009) 015,
  [\href{http://xxx.lanl.gov/abs/hep-ph/0905.0662}{{\tt hep-ph/0905.0662}}].

\bibitem{aristizabal09b}
D.~Aristizabal~Sierra, L.~A. Munoz, and E.~Nardi, {\it {Purely Flavored
  Leptogenesis}},  {\em Phys. Rev.} {\bf D80} (2009) 016007,
  [\href{http://xxx.lanl.gov/abs/hep-ph/0904.3043}{{\tt hep-ph/0904.3043}}].

\bibitem{gonzalezgarcia09}
M.~C. Gonzalez-Garcia, J.~Racker, and N.~Rius, {\it {Leptogenesis without
  violation of B-L}},  {\em JHEP} {\bf 11} (2009) 079,
  [\href{http://xxx.lanl.gov/abs/hep-ph/0909.3518}{{\tt hep-ph/0909.3518}}].

\bibitem{antusch09}
S.~Antusch, S.~Blanchet, M.~Blennow, and E.~Fernandez-Martinez, {\it
  {Non-unitary Leptonic Mixing and Leptogenesis}},  {\em JHEP} {\bf 01} (2010)
  017, [\href{http://xxx.lanl.gov/abs/hep-ph/0910.5957}{{\tt
  hep-ph/0910.5957}}].

\bibitem{grossman03}
Y.~Grossman, T.~Kashti, Y.~Nir, and E.~Roulet, {\it {Leptogenesis from
  Supersymmetry Breaking}},  {\em Phys. Rev. Lett.} {\bf 91} (2003) 251801,
  [\href{http://xxx.lanl.gov/abs/hep-ph/0307081}{{\tt hep-ph/0307081}}].

\bibitem{dambrosio03}
G.~D'Ambrosio, G.~F. Giudice, and M.~Raidal, {\it {Soft leptogenesis}},  {\em
  Phys. Lett.} {\bf B575} (2003) 75--84,
  [\href{http://xxx.lanl.gov/abs/hep-ph/0308031}{{\tt hep-ph/0308031}}].

\bibitem{fong08}
C.~S. Fong and M.~C. Gonzalez-Garcia, {\it {Flavoured Soft Leptogenesis}},
  {\em JHEP} {\bf 06} (2008) 076,
  [\href{http://xxx.lanl.gov/abs/hep-ph/0804.4471}{{\tt hep-ph/0804.4471}}].

\bibitem{pilaftsis08}
A.~Pilaftsis, {\it {Electroweak Resonant Leptogenesis in the Singlet Majoron
  Model}},  {\em Phys. Rev.} {\bf D78} (2008) 013008,
  [\href{http://xxx.lanl.gov/abs/hep-ph/0805.1677}{{\tt hep-ph/0805.1677}}].

\end{thebibliography}
\end{document}